\documentclass[10pt,conference]{IEEEtran}
\IEEEoverridecommandlockouts

\usepackage{cite}
\usepackage{amsmath,amssymb,amsfonts}
\usepackage{algorithm}
\usepackage{algorithmicx}
\usepackage{algpseudocode}  
\usepackage{graphicx}
\usepackage{textcomp}
\usepackage{booktabs}
\usepackage{xcolor}
\usepackage{subcaption}

\title{BucketServe: Bucket-Based Dynamic Batching for Smart and Efficient LLM Inference Serving}

\author{
    \IEEEauthorblockN{Wanyi Zheng\textsuperscript{1,2}, Minxian Xu\textsuperscript{2,*\thanks{*M. Xu is the corresponding author.}}, Shengye Song\textsuperscript{1,2}, Kejiang Ye\textsuperscript{2}}
    \IEEEauthorblockA{\textsuperscript{1}Southern University of Science and Technology}
    \IEEEauthorblockA{\textsuperscript{2}Shenzhen Institutes of Advanced Technology, Chinese Academy of Sciences}
    \IEEEauthorblockA{Email: \{wy.zheng, mx.xu, sy.song, kj.ye\}@siat.ac.cn}
}

\begin{document}

\maketitle

\begin{abstract}
Large language models (LLMs) have become increasingly popular in various areas, traditional business gradually shifting from rule-based systems to LLM-based solutions. However, the inference of LLMs is resource-intensive or latency-sensitive, posing significant challenges for serving systems. Existing LLM serving systems often use static or continuous batching strategies, which can lead to inefficient GPU memory utilization and increased latency, especially under heterogeneous workloads. These methods may also struggle to adapt to dynamic workload fluctuations, resulting in suboptimal throughput and potential service level objective (SLO) violations. In this paper, we introduce BucketServe, a bucket-based dynamic batching framework designed to optimize LLM inference performance. By grouping requests into size-homogeneous buckets based on sequence length, BucketServe minimizes padding overhead and optimizes GPU memory usage through real-time batch size adjustments preventing out-of-memory (OOM) errors. It introduces adaptive bucket splitting/merging and priority-aware scheduling to mitigate resource fragmentation and ensure SLO compliance. Experiment shows that BucketServe significantly outperforms UELLM in throughput, achieving up to 3.58× improvement. It can also handle 1.93× more request load under the SLO attainment of 80\% compared with DistServe and demonstrates 1.975× higher system load capacity compared to the UELLM.
\end{abstract}


    \section{Introduction}
    \label{sec:intro} In the past few years, with the advancement of the
    artificial intelligence (AI), large language models, such as GPT-4\cite{openai2024gpt4technicalreport,icapp2024}, Deepseek-R1\cite{deepseekai2025deepseekr1incentivizingreasoningcapability},
    and LLaMA\cite{grattafiori2024llama3herdmodels}, based on Transformer\cite{vaswani2023attentionneed} architecture and its variants (e.g., Encoder-Only
    and Decoder-Only) have demonstrated remarkable capabilities in natural language
    processing (NLP). Benefiting from the attention mechanism inherent in the
    Transformer architecture, LLMs can capture long-range dependencies and contextual information, enabling them to perform a wide range of tasks, such as text generation, translation, summarization, multi-turn dialogue and delivering precise automated question-answering services\cite{li2021pretrainedlanguagemodelstext,adams2023sparsedensegpt4summarization,yi2024surveyrecentadvancesllmbased}. Thereby expanding the practical applications of large-scale models. Leveraging their massive parameter scales and deep
    learning capabilities, these models effectively capture the complexity and semantic
    structures of human language, marking a significant step toward achieving
    general AI  and human capability.

    However, these advancements in LLMs have also driven a growing demand for high-performance
    inference systems capable of meeting the computational and memory requirements
    of increasingly LLMs~\cite{zhong2024distserve,kwon2023efficient,agrawal2023sarathiefficientllminference,he2024uellmunifiedefficientapproach,dao2023flashattention2,aminabadi2022deepspeedinferenceenablingefficient,li2024llminferenceservingsurvey}. Unlike traditional online services that rely
    on lightweight models or database queries\cite{zhong2024distserve}, LLM inference typically demands substantial
    GPU resources and storage capacity in AI infrastructure (e.g. cloud computing data centers). In particular, real-time services
    such as chatbots\cite{zheng2023judging}, virtual assistants, and interactive AI systems\cite{zhong2024distserve} require strict
    adherence to SLOs, where low latency and fast
    response times are essential to user satisfaction. Performance degradation for LLM inference services deployed on cloud can lead to poor user experience, reduced engagement, and potential
    business impact\cite{TSC2025Xu}\cite{TAAS2025Xu}. On the other hand, there are applications with more relaxed
    latency constraints, where throughput and resource utilization are the primary concerns.
    In these cases, the focus should shift toward maximizing system efficiency and
    minimizing cost per request. 
    
    Therefore, it is crucial to design flexible and scalable LLM inference systems that can adapt to diverse application requirements\cite{he2024uellmunifiedefficientapproach}.
    For time-sensitive tasks, optimization efforts should prioritize reducing end-to-end
    latency and ensuring SLO compliance. For less latency-critical workloads, the
    emphasis should be on improving throughput and optimizing resource allocation\cite{sun2024llumnix}.
    To address this challenge, researchers and engineers have explored various techniques
    to enhance the end-to-end performance of LLM inference, including dynamic
    batching\cite{Orca280922}, and efficient scheduling strategies\cite{sun2024llumnix}. Among them, recent research has shown increasing interest in disaggregated
    architectures\cite{zhong2024distserve,patel2024splitwiseefficientgenerativellm,hu2024memservecontextcachingdisaggregated}, such as DistServe\cite{zhong2024distserve}, which decouples the prefill and decoding phases
    to enable specialized optimizations for each stage. Building upon such
    architectural innovations, further improvements, especially in request scheduling, remain possible and highly valuable. Empirical studies from major cloud
    providers demonstrate that intelligent scheduling can significantly improve
    both throughput and resource efficiency without compromising service quality\cite{sun2024llumnix}.
    This makes advanced scheduling strategies an essential component in next-generation
    LLM serving systems, particularly under high concurrency and heterogeneous
    workload conditions.

    In this paper, we propose BucketServe, a bucket-based dynamic batching framework for efficient LLM inference, designed to achieve both high throughput and low latency under concurrent workloads. Built upon vLLM \cite{kwon2023efficient}, BucketServe extends its capabilities to support request scheduling in a disaggregated serving architecture. The core idea of BucketServe is to group incoming requests into buckets based on their sequence lengths, enabling fine-grained and adaptive batching decisions. This approach allows the system to prioritize latency-sensitive requests while maximizing overall throughput. Experimental results demonstrate that BucketServe significantly improves system efficiency and scalability compared to baselines while introducing negligible overhead. Our contributions can be summarized as follows:
    \begin{itemize}
        \item \textbf{Mitigate resource inefficiency in static batching
            strategies:} We address the critical issue of resource waste caused by naive batching approaches through request orchestration based on input characteristics (e.g., sequence length and memory footprint). 

        \item \textbf{Introduce a bucket-based dynamic batching strategy:} We propose a novel algorithm that dynamically adjusts the border of the buckets and the batch size according to real-time GPU memory constraints and workloads while prioritizing high-priority requests.

        \item \textbf{Demonstrate the superior performance of BucketServe:} We comprehensively evaluate BucketServe on representative LLM serving workloads, showing its effectiveness in improving throughput and handling high-concurrency scenarios. Our results indicate that BucketServe significantly reduces SLO violations under high heterogeneous workloads compared to existing systems.
    \end{itemize}
\section{Background and Motivation}
    \label{sec:background}

    As LLMs are increasingly integrated into applications such as text generation and conversational AI, LLM-based services have rapidly transitioned from research prototypes to large-scale production systems. The dominant deployment paradigm is the client-server architecture, where user requests are funneled through a gateway to backend inference servers responsible for executing generative AI tasks and returning results. These services encompass both real-time interactive scenarios and offline batch processing tasks. 

    Despite their versatility, LLM inference remains highly resource-intensive: a single inference with a large model can occupy several gigabytes of GPU memory, and the autoregressive generation process leads to considerable variability in response latency, varying from seconds to minutes. These factors pose significant challenges for maintaining SLOs under high concurrency. 
    \begin{figure}[]
        \centering
        \includegraphics[width=0.98\columnwidth]{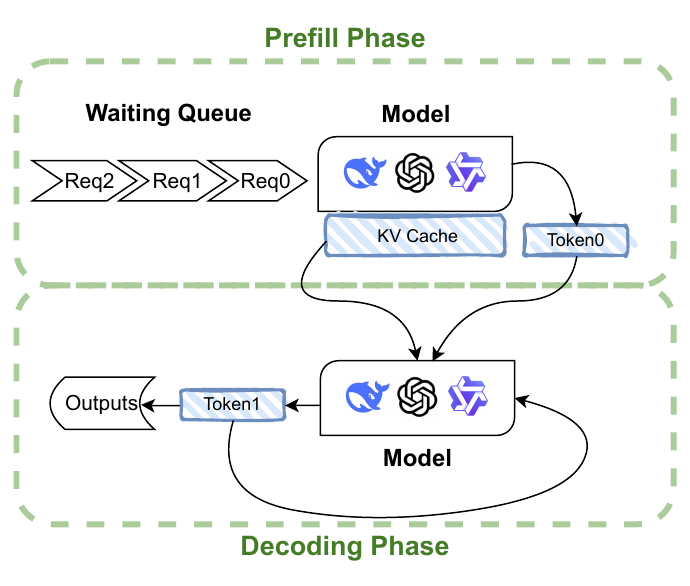}
        \caption{Prefill and Decoding disaggregation architecture for LLM inference. }
        \label{fig:disaggregated-architecture}
    \end{figure}
    \subsection{Challenges in Disaggregated Architecture}

    Disaggregating LLM inference into prefill and decoding phases is illustrated in Fig.~\ref{fig:disaggregated-architecture}, where the LLM reads the entire prompt at once and computes the Key-Value (KV) Cache in the prefill phase. During decoding phase, the model generates new tokens incrementally, one token at a time, based on the existing context and the previously computed KV cache. It continuously updates the KV cache and passes the newly generated token back to the model. The Disaggregation enables specialized optimizations for each stage, but also introduces several significant challenges that must be addressed to achieve efficient and reliable serving.

    \textbf{1) Resource Contention and Imbalance:} The prefill phase is highly compute-intensive, requiring substantial GPU computation to process input prompts in parallel and construct the initial KV cache. In contrast, the decoding phase generates output tokens sequentially and is often bottlenecked by memory bandwidth due to frequent KV cache accesses. This heterogeneity can lead to resource imbalance, where compute resources may be under-utilized during decoding, while memory bandwidth becomes a limiting factor, resulting in suboptimal overall throughput.

    \textbf{2) Scheduling Complexity:} Disaggregated architectures require two-stage scheduling: requests must be efficiently orchestrated for both prefill and decoding phases. This increases scheduling complexity, as requests may experience head-of-line blocking if long-sequence prefill jobs delay shorter ones, or if decoding jobs accumulate and saturate memory bandwidth. Moreover, scheduling strategies must account for request-specific attributes such as sequence length, priority, and latency sensitivity to avoid resource monopolization and ensure SLO compliance. Coordinating scheduling decisions across both phases is non-trivial, especially under heterogeneous and bursty workloads.

    \textbf{3) Batching Challenges:} Batching is essential for maximizing GPU utilization, but in disaggregated architectures, the optimal batching strategy may differ between prefill and decoding. Grouping requests with diverse sequence lengths in a batch leads to excessive padding, wasting GPU memory and compute. The prefill phase benefits from large, homogeneous batches to maximize parallelism, while the decoding phase, due to its sequential nature and variable output lengths, may require smaller or more dynamic batches. Designing batching strategies that adapt to the requirements of both phases, minimize padding, and avoid fragmentation is a key challenge.

    \textbf{4) Coordination and Data Transfer Overhead:} The separation of prefill and decoding phases necessitates transferring large KV cache between them, often across GPUs or nodes. This introduces additional coordination and communication overhead, which can degrade latency and throughput if not carefully managed.

    For the first and fourth challenges, which are highly dependent on hardware characteristics, we adopt the best configurations recommended by prior work~\cite{zhong2024distserve}. The remaining challenges require advanced scheduling and batching mechanisms that are aware of both request characteristics and system resource constraints. 

    \subsection{Opportunities}

\begin{figure}
  \centering
  \begin{subfigure}{0.9\linewidth}
    \includegraphics[width=\linewidth]{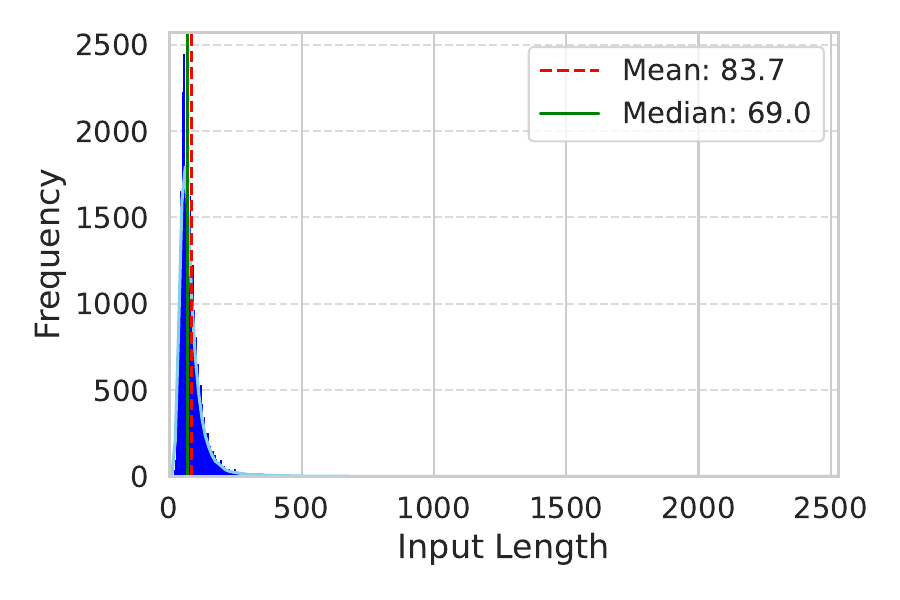}
    \caption{Distribution of Alpaca.}
    \label{fig:alpaca-distribution}
  \end{subfigure}


  \begin{subfigure}{0.9\linewidth}
    \includegraphics[width=\linewidth]{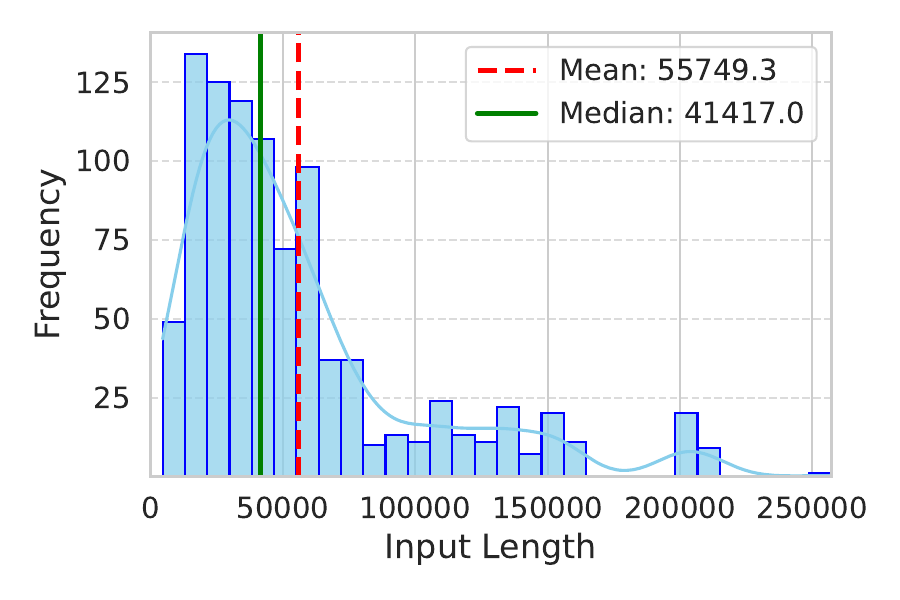}
    \caption{Distribution of LongBench.}
    \label{fig:longserve-distribution}
  \end{subfigure}

  \caption{Distribution of LLM Requests.}
  \label{fig:request-len}
\end{figure}

\begin{figure}
  \centering
  \begin{subfigure}[b]{0.9\linewidth}
    \includegraphics[width=\linewidth]{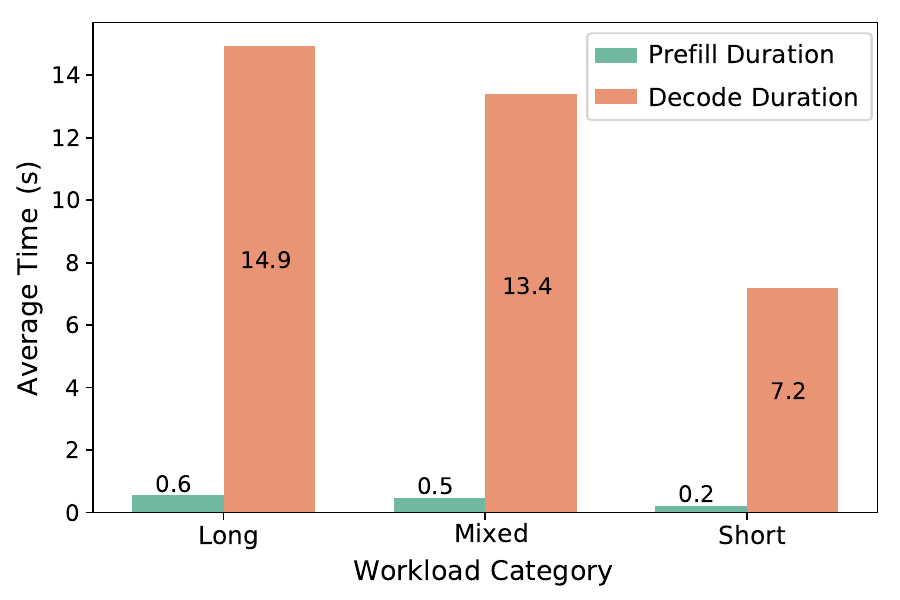}
    \caption{Average Batch Execution Cost.}
    \label{fig:m1}
  \end{subfigure}
  \vspace{0.3cm}
  \begin{subfigure}[b]{0.9\linewidth}
    \includegraphics[width=\linewidth]{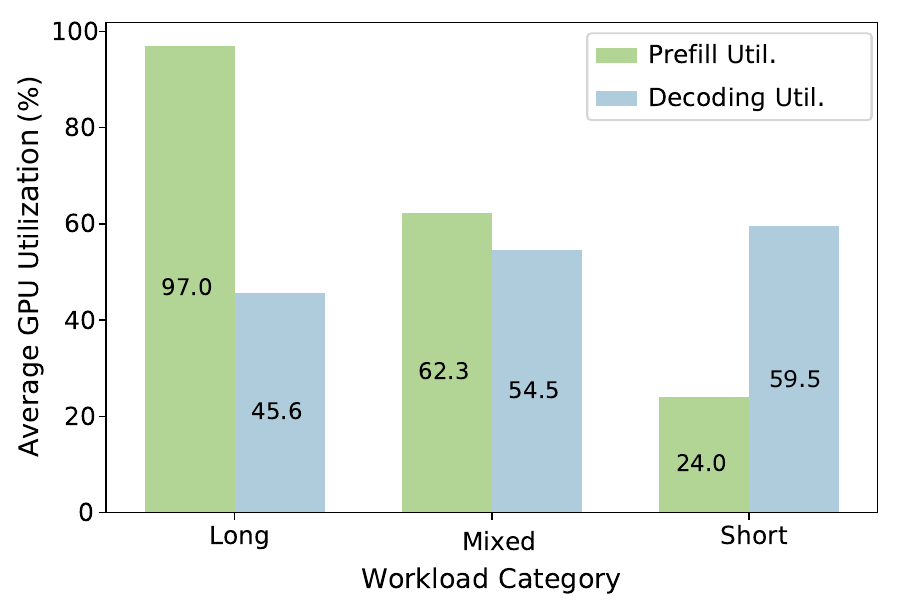}
    \caption{Average GPU Utilization.}
    \label{fig:m2}
  \end{subfigure}
  \caption{Performance of batch execution and GPU utilization across different workload types. Long refers to those exceeding 1024 tokens sampled from LongBench, Short denotes those under 256 tokens from Alpaca, and Mixed contains sequences from both datasets following a long-tail distribution pattern.}
  \label{fig:Motivation}
\end{figure}

As shown in Figs.~\ref{fig:request-len} and \ref{fig:Motivation}, here we use Stanford Alpaca\cite{alpaca} and LongBench\cite{bai2024longbench} datasets to conduct the case study. Our analysis of request characteristics and phase-specific resource utilization reveals two critical opportunities for optimization:

\textbf{Dynamic Batching with Prefill Bucketing.}  
The heterogeneous sequence length distribution (Fig.~\ref{fig:request-len}) shows that the input lengths exhibit a heterogeneous distribution, with Alpaca sequences averaging~83 tokens and LongBench sequences showing a long-tail pattern (median~41,417 tokens). For LongBench's ultra-long sequences, we truncate them to the model and the dominance of long sequences in execution time (Fig.~\ref{fig:m1}) motivate a two-stage batching strategy. For the prefill phase, we adopt bucketing-based batching: requests are grouped into size-homogeneous buckets (e.g., [0–256 tokens], [256–1024 tokens]) to minimize padding overhead. This allows direct batch formation from buckets with predictable memory requirements, ensuring efficient GPU utilization for static input sequences. In contrast, for the decoding phase that have output lengths, we apply continuous batching \cite{Orca280922} to dynamically incorporate newly arriving tokens without waiting for full batches. This combination leverages the static nature of prefill inputs while adapting to the dynamic decoding process.

\textbf{Bucket-Aware Scheduling.}  
The low GPU utilization observed in mixed-batch (containing samples from two datasets) decoding (Fig.~\ref{fig:m2}) highlights the need for scheduling strategies that align with workload characteristics. By first partitioning requests into buckets with minimal overhead, then applies phase-aware dispatching tailored to specific performance goals:

In scenarios where requests per second (RPS) is the priority, we consider applying shortest-job-first (SJF) within buckets to minimize queuing latency and maximize throughput. By prioritizing shorter requests, this strategy ensures timely responses for latency-sensitive tasks, even at the cost of reduced GPU memory utilization for long sequences.

For workloads prioritizing token-per-second throughput, we can use longest-job-first (LJF) within buckets to group long sequences. This maximizes GPU utilization during decoding by leveraging parallelized attention operations and reducing padding overhead, albeit at the expense of increased latency for individual requests.

We address these issues by proposing adaptive bucketing and dynamic batching strategies proposed above that jointly optimize resource utilization, scheduling, and memory efficiency across both phases of disaggregated LLM inference. The implementation details are presented in Section~\ref{sec:core-algorithm}.


 \section{System Overview}
\label{sec:system-overview}

\begin{figure}[t]
    \centering
    \includegraphics[width=0.98\columnwidth]{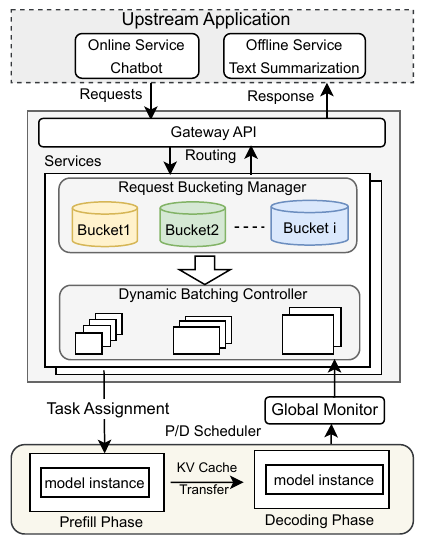}
    \caption{The Architecture of BucketServe.}
    \label{fig:system-overview}
\end{figure}

As illustrated in Fig.~\ref{fig:system-overview}, BucketServe adopts a three-tier architecture, decoupling application workloads from hardware resources via a middleware layer. The top-tier application layer supports diverse LLM services, including both online and offline tasks. When incoming requests arrive at the gateway, they are routed to the appropriate services based on their type and priority. The middleware layer then performs adaptive request bucketing, grouping requests into buckets according to their sequence length and task category.

The system employs a dynamic bucketing strategy to optimize resource utilization across varying workloads:

\textbf{Low-Load Scenarios}: In situations where the RPS is low and all requests can be processed in a single batch, the system consolidates all requests into a single bucket to minimize scheduling overhead.
   
\textbf{High-Load Scenarios}: If the number of requests in a bucket exceeds a threshold and exhibits uneven distribution (e.g., a mix of short and long sequences), the system dynamically adjusts the bucket count and boundaries to reduce padding overhead and improve batch efficiency. This adaptive mechanism ensures optimal performance across diverse workloads and is discussed in detail in Section~\ref{sec:core-algorithm}.

\textbf{Request Bucketing Manager} is responsible for grouping incoming requests into buckets based on their sequence lengths and task categories. It dynamically adjusts the number of buckets and their boundaries to minimize padding overhead and ensure efficient batching. In low-load scenarios, it consolidates all requests into a single bucket to reduce scheduling overhead. In high-load scenarios, it splits buckets when the number of requests exceeds a threshold or when the distribution of sequence lengths becomes uneven. This adaptive approach helps maintain optimal resource utilization and throughput across diverse workloads.

\textbf{Dynamic Batching Controller} takes requests from the buckets and groups them into batches for processing. It dynamically computes the optimal batch size based on current GPU memory constraints and bucket boundaries. By adjusting the batch size according to available resources, the controller prevents OOM errors while maximizing throughput. Additionally, the controller prioritizes requests that have been waiting the longest, ensuring they are processed quickly and efficiently.

\textbf{P/D Scheduler} manages the transition of requests between the prefill phase and the decoding phase. Once the requests are batched, they are assigned to the waiting queue of the prefill phase. A First-Come-First-Served (FCFS) scheduling strategy is adopted by the workers to process the requests in order of their arrival. This ensures fairness and simplicity in execution order. After the prefill phase completes, the corresponding KV cache is transferred to the decoding phase via NVLink, enabling high-bandwidth, low-latency communication between GPUs. In the decoding phase, continuous batching is applied, as the output length during token generation is typically variable and hard to predict.

\textbf{Global Monitor} continuously collects and aggregates system-wide metrics, including GPU memory usage, queue lengths, request arrival rates, average sequence length, and batch processing latency. These metrics provide a real-time view of resource availability and workload characteristics, enabling the system to make data-driven decisions.

In particular, the Global Monitor feeds critical information to the Dynamic Batching Controller and P/D Scheduler. For example, it informs the batching controller about current memory pressure and expected future demand, allowing it to adjust batch sizes dynamically without risking OOM errors. It also provides the scheduler with queue occupancy and waiting time statistics, supporting fair and efficient scheduling policies.

    \section{Adaptive Bucketing and Dynamic Batching with Memory Safety}
\label{sec:core-algorithm}

In this section, we present the design and implementation of our strategy. For offline tasks with relaxed latency constraints, requests are grouped into buckets based on sequence length to minimize padding overhead. For online tasks requiring low latency, the number of buckets and their boundaries are dynamically optimized in real-time. The batching module computes optimal batch sizes using the following equations to maximize throughput while preventing OOM errors.

The KV cache memory footprint for a batch is calculated as:
\begin{equation}
\begin{split}
\mathrm{Memory}_{\mathrm{KV\ cache}} = & \, 2 \times L \times H \times D \times \\
& S_{\mathrm{max}} \times B \times N,
\end{split}
\label{eq:kv_cache}
\end{equation}
where $L$ is the number of layers, $H$ is the number of attention heads, $D$ is the dimension per head, $S_{\mathrm{max}}$ is the maximum sequence length in the batch, $B$ is the bytes per element (e.g., 2 for FP16), and $N$ is the batch size.

To quantify the inefficiency introduced by padding, we define the \textbf{wasted memory ratio} as:
\begin{equation}
\mathrm{Waste}_{\mathrm{Ratio}} = \frac{S_{\mathrm{max}} - S_{\mathrm{avg}}}{S_{\mathrm{max}}},
\label{eq:waste_ratio}
\end{equation}
where $S_{\mathrm{avg}}$ is the average sequence length in the batch. Bucketing reduces $S_{\mathrm{max}} - S_{\mathrm{avg}}$, thereby minimizing memory waste.

To further analyze the expected memory waste across all buckets, we define the \textbf{expected waste rate} as:
\begin{equation}
\mathbb{E}[\text{Waste}] = \sum_{b=1}^{K} \int_{L_b}^{U_b} \left(1 - \frac{S}{U_b}\right) f(S) dS,
\label{eq:expected_waste}
\end{equation}
where $f(S)$ is the distribution of incoming request lengths of batch $S$, and $[L_b, U_b)$ denotes the range of the $b$-th bucket. This formulation provides a more comprehensive view of how different bucketing strategies affect overall padding overhead.

The \textbf{optimal bucket boundary} that minimizes $\mathbb{E}[\text{Waste}]$ is given by:
\begin{equation}
U_b^* = \frac{\int_{L_b}^{U_b} S f(S) dS}{\int_{L_b}^{U_b} f(S) dS}.
\label{eq:optimal_boundary}
\end{equation}
This indicates that the upper bound of each bucket should be set to the \textbf{conditional expectation} of sequence lengths within that bucket. Our adaptive algorithm approximates this condition dynamically during runtime. 

Although the optimal bucket boundary that minimizes \textbf{E}[\text{Waste}] is theoretically defined as the conditional expectation of sequence lengths within a bucket, it is computationally expensive to calculate in practice. Moreover, since request length distributions can change over time, maintaining such boundaries dynamically introduces significant overhead and algorithmic complexity. To address this challenge, we adopt a simple but efficient approach based on interval bisection, which approximates the optimal boundary while keeping the system lightweight and responsive. 

To ensure memory safety, the scheduler reserves 10\% of GPU memory for system overheads. The \textbf{safe available memory} is computed as:
\begin{equation}
M_{\mathrm{safe}} = 0.9 \times M_{\mathrm{remain}},
\label{eq:safe_memory}
\end{equation}
where $M_{\mathrm{remain}}$ is the remaining memory after model allocations.

The \textbf{maximum safe batch size} is determined as:
\begin{equation}
\begin{split}
N_{\max} = \max \biggl\{ N \in \mathbb{N} \, \biggm| \, \sum_{i=1}^{N} S_{i} \leq 
\frac{M_{\mathrm{safe}}}{2 L H D B} \biggr\},
\end{split}
\label{eq:max_batch_size}
\end{equation}
which ensures that the KV cache remains within safe limits.

\begin{algorithm}[t]
\caption{Adaptive Bucketing Mechanism}
\label{alg:bucket-management-only}
\begin{algorithmic}[1]
\Require Buckets $B$, maximum length $L_{\max}$ 
\Require Minimum split size $m = N_{\max}$, threshold $\theta = 0.5$
\State Initialize $B \gets \{ [0, L_{\max}) \}$
\For{each request $r = (S, t, \texttt{type})$}
    \For{each bucket $b \in B$}
        \If{$b_{\text{low}} \leq S < b_{\text{up}}$}
            \State Add $r$ to $b.\texttt{requests}$
            \State \textbf{break}
        \EndIf
    \EndFor
\EndFor
\Statex
\Function{AdjustBuckets}{}
    \State $\texttt{total} \gets \sum_{b} |b.\texttt{requests}|$
    \If{$\texttt{total} < N_{\max}$}
        \State $B \gets \{ [0, L_{\max}) \}$
    \Else
        \State $\texttt{split\_list} \gets \emptyset$
        \For{each bucket $b \in B$}
            \State $m_b \gets (b_{\text{low}} + b_{\text{up}})/2$
            \State $C_s \gets |\{ r \in b.\texttt{requests} \mid r.S < m_b \}|$
            \If{$C_s/|b.\texttt{requests}| > \theta$ \textbf{and} $|b.\texttt{requests}| > m$}
                \State $\texttt{split\_list} \gets \texttt{split\_list} \cup \{b\}$
            \EndIf
        \EndFor
        \For{each $b \in \texttt{split\_list}$}
            \State $\text{mid} \gets (b_{\text{low}} + b_{\text{up}})/2$
            \State Create $b_L \gets [b_{\text{low}}, \text{mid})$
            \State Create $b_R \gets [\text{mid}, b_{\text{up}})$
            \State Partition $b.\texttt{requests}$ into $b_L$, $b_R$ by $S$
            \State $B \gets B \cup \{b_L, b_R\} \setminus \{b\}$
        \EndFor
    \EndIf
\EndFunction
\end{algorithmic}
\end{algorithm}
Algorithm~\ref{alg:bucket-management-only} implements an adaptive bucketing mechanism to manage request scheduling and resource allocation. Initially, the system starts with a single bucket covering the full sequence length range $[0, L_{\max})$. Requests are assigned to the appropriate bucket based on their sequence length $S$ (lines 2-6), where each incoming request is matched against the current set of buckets.

When the total number of requests across all buckets exceeds $N_{\max}$ (computed via Eq.~(\ref{eq:max_batch_size})), the algorithm triggers a splitting strategy. Specifically, for each bucket $b$, if over 50\% of its requests have sequence lengths below the midpoint $m_b = (b_{\text{low}} + b_{\text{up}})/2$ and the bucket contains more than $N_{\max}$ requests, it is split into two sub-buckets. This midpoint-based decision (lines 14-19), serves as an approximation to the optimal boundary defined in Eq.~(\ref{eq:optimal_boundary}). Splitting helps reduce padding by grouping similar-length sequences together. Then it conducts a partitioning of requests into the new buckets based on their sequence lengths (lines 23-29). This process continues until all buckets are split depending on the current workload.

Conversely, when the total number of requests drops below $N_{\max}$, all buckets merge back into a single one to minimize scheduling overhead (lines 11-13). After bucketing, offline tasks use SJF or LJF scheduling within buckets to optimize throughput, while online tasks prioritize buckets based on earliest request arrival time to meet SLOs. The algorithm balances efficiency and fairness through dynamic adjustments, leveraging thresholds like $\theta = 0.5$ for splitting decisions and $N_{\max}$ for batch size constraints. 

The potential performance improvements include optimizing bucket search with data structures such as binary trees and refining splitting criteria using distribution-aware methods.


\textbf{Algorithm Complexity Analysis.} The adaptive bucketing algorithm consists of two main components. The first is assigning requests to buckets: for each of the $n$ requests, the algorithm traverses all $k$ buckets to find the appropriate one, resulting in a time complexity of $O(n \cdot k)$. The second component is the \texttt{AdjustBuckets} function. Calculating the total number of requests requires $O(k)$ time. For each bucket, the algorithm performs constant-time computations and checks, also totaling $O(k)$. Splitting buckets and partitioning requests can be done in $O(k)$ time overall. Therefore, the time complexity of the adjustment step is $O(k)$. In summary, the overall time complexity of the adaptive bucketing mechanism is $O(n \cdot k + 4k)$.

 \section{Performance Evaluations}
    In this section, we evaluate the performance of BucketServe compared with two
    baseline systems. Because BucketServe is originally designed for offline tasks which have high throughput requirement, we adapted it to online tasks, we found that the performance of BucketServe outperforms baseline systems in high concurrency scenarios.

    \label{sec:evaluation}
    \subsection{Experiments Setup}
    \textbf{Testbed.} We evaluated BucketServe on a single-node cluster equipped with 4 NVIDIA A100 40GB GPUs interconnected via NVLink for high-bandwidth GPU-to-GPU communication. The node also features 64 CPU cores and 1TB of high-speed SSD storage. Our serving framework is built on the vLLM backend, which supports continuous batching and paged attention for high-performance inference.
    
    \textbf{Models and Datasets.} We selected the LLaMA-2\cite{touvron2023llama2openfoundation} and OPT\cite{zhang2022optopenpretrainedtransformer} series as representative foundation models due to their standardized architectures and widespread adoption in academic research. To evaluate performance across diverse task profiles, we used two benchmarking datasets: Stanford Alpaca\cite{alpaca}, characterized by short sequence lengths, and LongBench\cite{bai2024longbench}, designed for long-text summarization scenarios. The Mixed configuration refers to a hybrid of these two datasets. This setup enables a comprehensive evaluation spanning both short and long sequence processing paradigms.
    


    \textbf{Metrics.} BucketServe's performance evaluation employs two distinct metric categories tailored to task types. For offline workloads, we prioritize throughput (measured in tokens per second) and pay attention to resource utilization to quantify system efficiency in handling large-scale request volumes. In online scenarios, we adopt SLOs attainment rate and service load capacity as the primary metrics.
  \begin{figure*}[htbp]
  \centering

  \begin{subfigure}[b]{0.32\linewidth}
    \includegraphics[width=\linewidth]{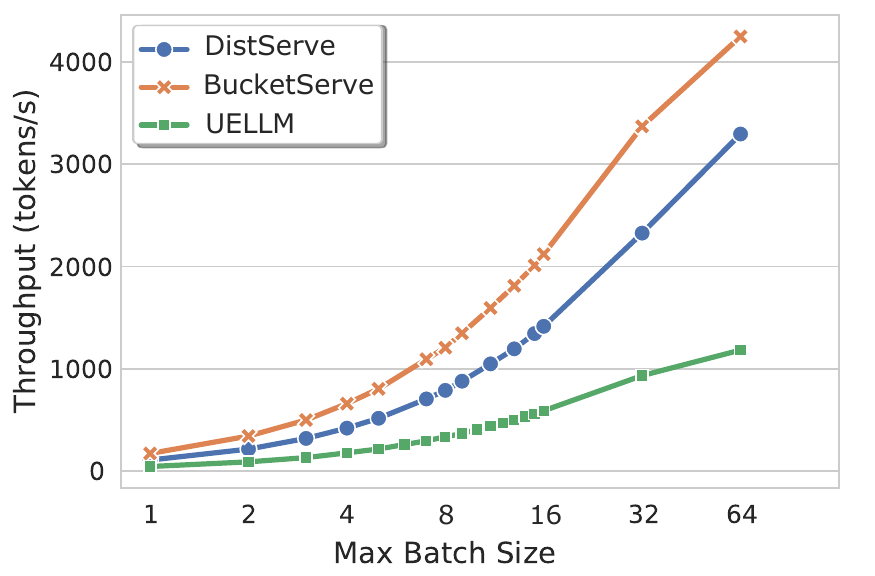}
    \caption{Throughput under Mixed Workload.}
    \label{fig:offline-performance}
  \end{subfigure}
  \hfill
  \begin{subfigure}[b]{0.32\linewidth}
    \includegraphics[width=\linewidth]{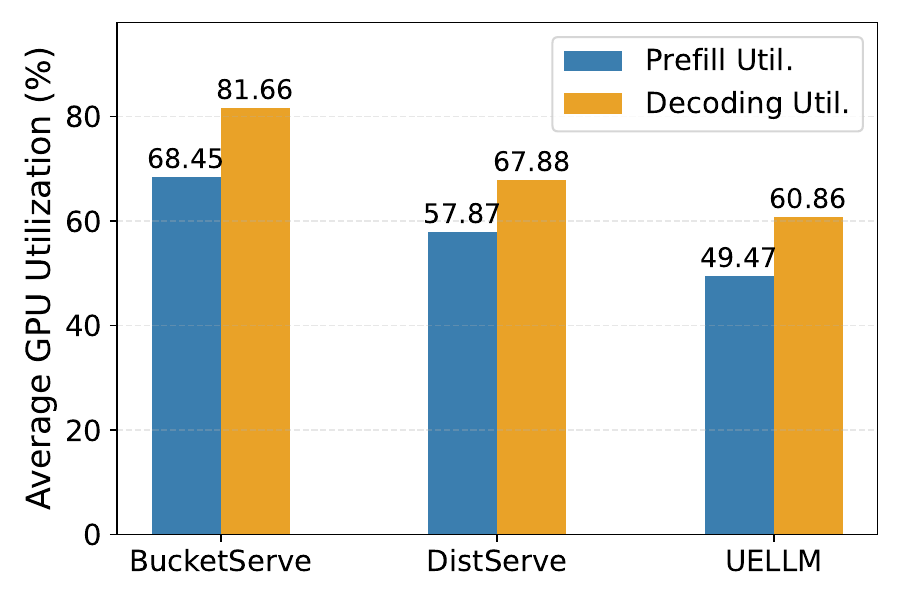}
    \caption{Average GPU Utilization.}
    \label{fig:gpu-util}
  \end{subfigure}
  \hfill
  \begin{subfigure}[b]{0.32\linewidth}
    \includegraphics[width=\linewidth]{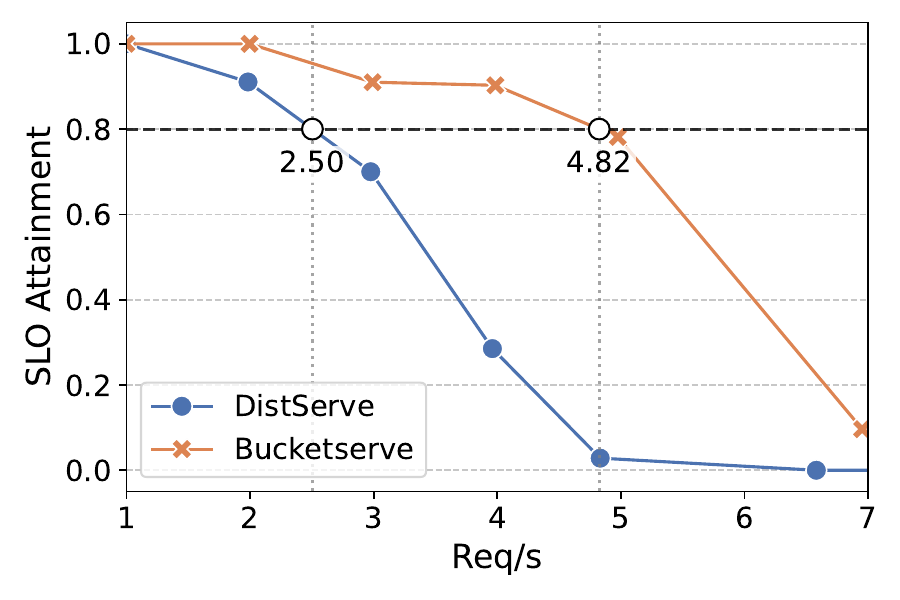}
    \caption{SLO Attainment of Alpaca.}
    \label{fig:alpaca}
  \end{subfigure}

  \vspace{0.3cm} 

  \begin{subfigure}[b]{0.32\linewidth}
    \includegraphics[width=\linewidth]{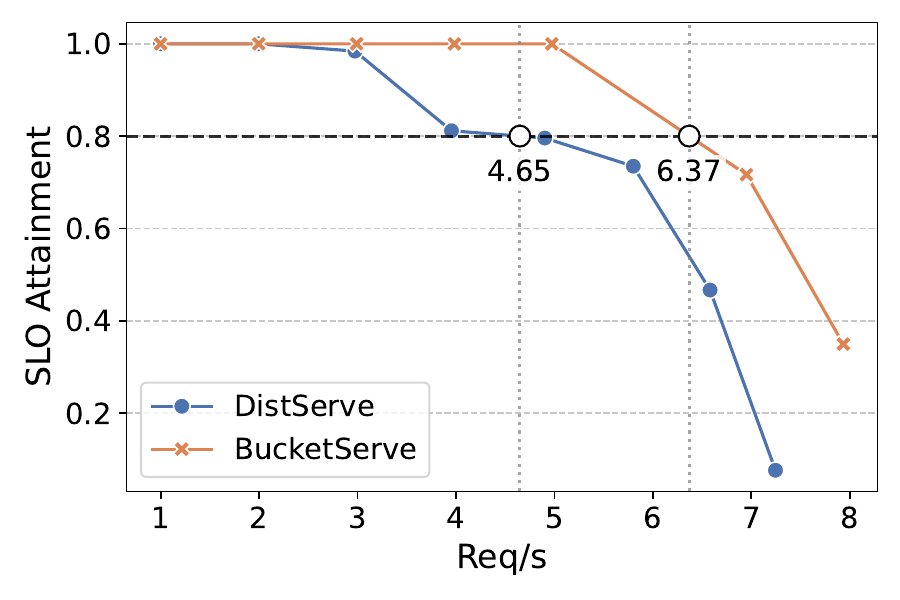}
    \caption{SLO Attainment of Mixed.}
    \label{fig:Longbench}
  \end{subfigure}
  \hfill
  \begin{subfigure}[b]{0.32\linewidth}
    \includegraphics[width=\linewidth]{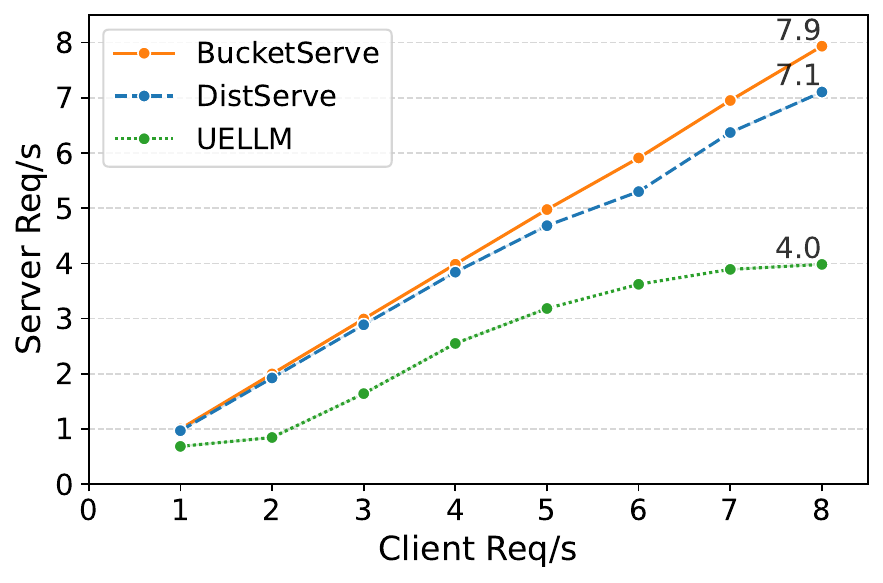}
    \caption{RPS under Alpaca.}
    \label{fig:rps-alpaca}
  \end{subfigure}
  \hfill
  \begin{subfigure}[b]{0.32\linewidth}
    \includegraphics[width=\linewidth]{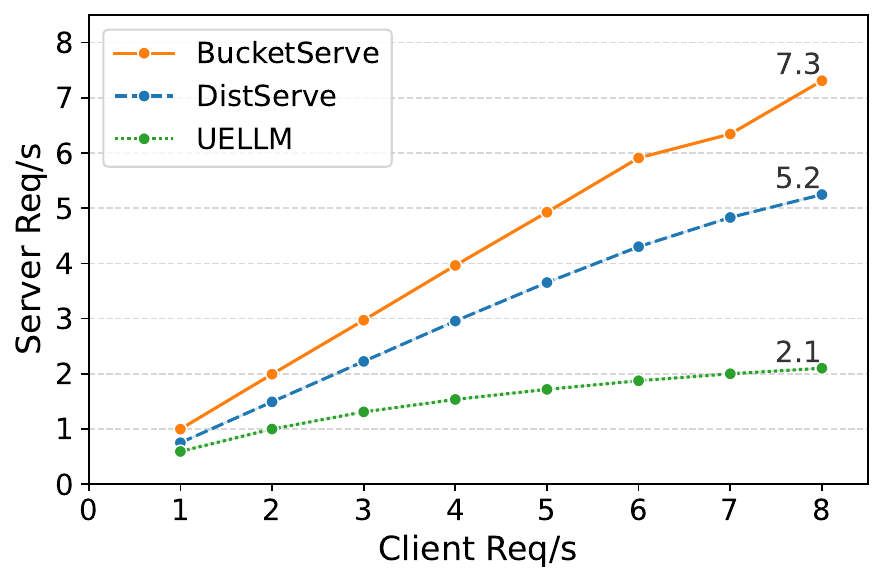}
    \caption{Server RPS under Mixed.}
    \label{fig:rps-longbench}
  \end{subfigure}

  \caption{End-to-End performance evaluation across offline and online dimensions.}
  \label{fig:comprehensive-performance}
\end{figure*}

    \textbf{Baselines.} We compare BucketServe with two state-of-the-art baseline systems including both aggregated and disaggregated architectures:
    \begin{itemize}
        
        \item \textbf{UELLM}\cite{he2024uellmunifiedefficientapproach}: A unified LLM serving framework that integrates resource profiling, batch scheduling, and deployment optimization. It employs a fine-tuned LLM to predict resource demands, batches queries based on predicted profiles, and deploys models considering hardware topology. However, UELLM still couples prefill/decoding phases and lacks dynamic adaptation to workload fluctuations, leading to suboptimal latency-SLO trade-offs under heterogeneous traffic patterns.
       
       \item  \textbf{DistServe}\cite{zhong2024distserve}: A disaggregated serving system that decouples prefill and decoding computation to eliminate phase interference. It co-optimizes resource allocation for each phase and minimizes communication overhead via bandwidth aware placement. While under high concurrency, DistServe lacks specialized process leading to suboptimal performance in heterogeneous workloads.
    \end{itemize}
  
    \subsection{Experiments Analysis}
    \label{sec:end-to-end-performance}
    We first evaluate the performance of BucketServe in offline tasks with those two baselines. To study the throughput and average resource utilization, we conducted experiments using Llama2-13B with varying batch sizes and sequence lengths. We randomly sampled requests from the Alpaca and LongBench datasets to simulate real-world workloads. The results are shown in Fig.~\ref{fig:offline-performance} and Fig.~\ref{fig:gpu-util}. The max batch size represents the largest number of requests the system can actually handle in one batch under the current setup, reflecting its realizable capacity. From this figure, we can see that BucketServe achieves the highest throughput and average GPU utilization especially under more requests. Specifically, BucketServe outperforms UELLM by 3.58× and DistServe 1.31× in throughput under high workloads, While dynamic batching improve average GPU utilization to 81.66\%. This demonstrates the effectiveness of our adaptive bucketing and dynamic batching strategy in optimizing resource utilization and throughput.

    For online tasks, we conducted experiments under varying client RPS and SLO requirements across different datasets. As shown in Fig.~\ref{fig:alpaca} and Fig.~\ref{fig:Longbench}, we compared BucketServe with DistServe using two datasets. We measured the server RPS and analyzed its relationship with SLO attainment.

    As the server RPS increases, the SLO attainment of both systems decreases. At an SLO attainment level of 80\%, BucketServe achieves 1.37× and 1.93× higher server RPS than DistServe on the Alpaca and Mixed datasets, respectively. This indicates that our system performs better under high workloads and for heterogeneous length requests.

    We also studied the relationship between the client request sending rate and the server processing rate, as it reflects the system's capability to handle high concurrency. From Fig.~\ref{fig:rps-alpaca} and Fig.~\ref{fig:rps-longbench}, we observe that, under both datasets, BucketServe’s server RPS closely follows the ideal y = x line, indicating excellent scalability and efficiency.

    On the Alpaca dataset, DistServe shows little performance degradation, but BucketServe still outperforms UELLM by 1.975× in terms of server RPS. On the Mixed dataset, BucketServe exhibits almost no performance degradation and achieves 1.4× and 3.47× higher server RPS than DistServe and UELLM, respectively. These results demonstrate BucketServe’s superior effectiveness, particularly under high workloads and for long-sequence processing.

\begin{figure}
  \centering
  \begin{subfigure}{\linewidth}
    \includegraphics[width=\linewidth]{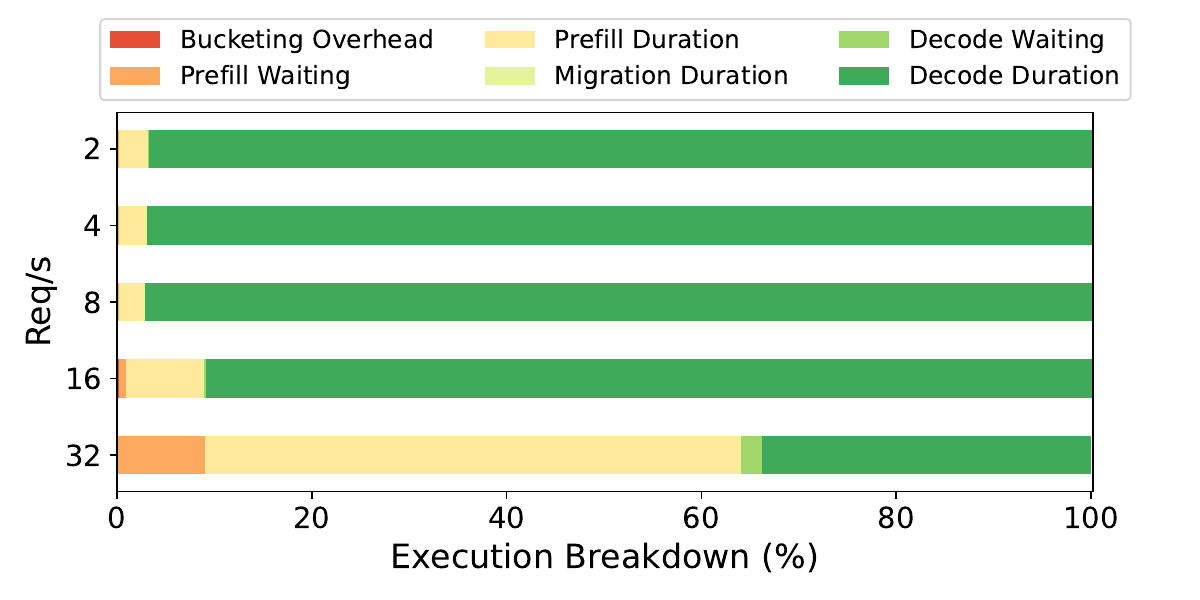}
    \caption{Execution Duration Breakdown.}
    \label{fig:a}
  \end{subfigure}


  \begin{subfigure}{\linewidth}
    \includegraphics[width=\linewidth]{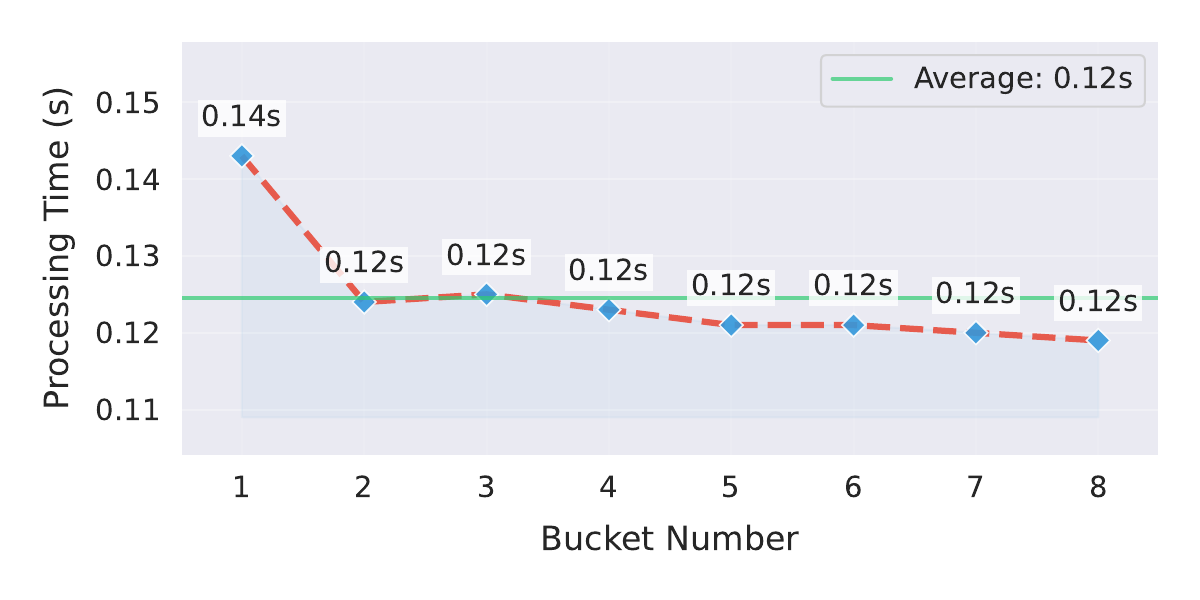}
    \caption{Distribution of Mixed.}
    \label{fig:b}
  \end{subfigure}

  \caption{End-to-End Latency Breakdown and Bucketing Overhead Analysis.}
  \label{fig:overhead}
\end{figure}

    Fig.~\ref{fig:overhead} presents a breakdown of the end-to-end execution duration for BucketServe and the associated bucketing overhead. As shown in Fig.~\ref{fig:a}, the decoding phase constitutes the majority of the execution time, accounting for approximately 90\% of the total under typical workloads. When the RPS is set to 32, some requests are queued waiting for the prefill stage, resulting in an increased average prefill duration.

    The red bar representing bucketing overhead is barely visible in the figure, indicating that the cost of bucketing is relatively low. As a result, our system introduces negligible overhead from bucketing and dynamic batching—less than 1\% of the total execution time. This highlights the efficiency of our adaptive bucketing and dynamic batching strategy in optimizing resource utilization and improving throughput.

    Fig.~\ref{fig:b} further illustrates the bucketing overhead. It shows that as the number of buckets increases, the algorithmic overhead remains stable, demonstrating its scalability and computational efficiency.

    To summarize, by introducing bucketing and dynamic batching, BucketServe effectively addresses the challenges of resource under-utilization and scheduling complexity in disaggregated LLM inference architectures, while introducing negligible overhead and maintaining performance comparable to prior approaches.

    \section{Related Work}
    \label{sec:related-work}
    Recent advancements in LLM inference optimization focus on enhancing end-to-end performance through diverse approaches\cite{zhong2024distserve,kwon2023efficient,agrawal2023sarathiefficientllminference,he2024uellmunifiedefficientapproach,aminabadi2022deepspeedinferenceenablingefficient,dao2022flashattention,dao2023flashattention2,patel2024splitwiseefficientgenerativellm,hu2024inferenceinterferencedisaggregatellm,hu2024memservecontextcachingdisaggregated,sun2024llumnix}, which can be broadly categorized into three domains: prefill-decoding disaggregation architectures, attention mechanism optimizations, and dynamic batching techniques.

    \textbf{Prefill-Decoding Disaggregation.} This optimization technique aims to decouple the distinct computational phases of LLM inference—Prefill and Decoding—to improve resource allocation. DistServe\cite{zhong2024distserve} pioneers this approach introducing an optimal configuration algorithm to dynamically allocate GPU resources and model parallelism strategies. In its design, to prioritize Time-to-First-Token (TTFT), waiting Prefill requests can preempt Decoding resources. While this reduces TTFT, it increases tail latency for Time-Between-Tokens (TBT) by up to 30\% due to frequent context switching. SplitWise\cite{patel2024splitwiseefficientgenerativellm} introduces a Prompt Phase and Token-Generation Phase, decomposing tasks into subtasks and mapping them to heterogeneous hardware via layer-wise KV cache transfer. This approach achieves 40\% higher resource utilization through task parallelism but incurs overhead in short-sequence workloads, increasing end-to-end latency. TetriInfer\cite{hu2024inferenceinterferencedisaggregatellm} partitions prompts into fixed-size chunks and employs a two-level scheduler to optimize batching, reducing average TTFT by 97\% and JCT by 47\% while cutting resource usage by 38\%. However, its chunked computation introduces communication overhead in high-throughput, long-sequence scenarios. Mooncake\cite{qin2024mooncakekvcachecentricdisaggregatedarchitecture} decouples KV cache storage from computation, leveraging Chunked Pipeline Parallelism (a variant of TeraPipe\cite{li2021terapipetokenlevelpipelineparallelism}) and prediction-guided early rejection to improve request throughput by 75\% under high load. However, inaccuracies in prediction models ($>$15\% error rates) lead to false rejections, harming throughput. LoongServe\cite{wu2024loongserveefficientlyservinglongcontext} introduces Elastic Sequence Parallelism, dynamically adjusting model parallelism degrees and optimizing KV cache migration to achieve higher throughput, while consistency maintenance for long sequences (1000+ tokens) adds significant overhead (e.g. 18\% computation). BucketServe also applies this architecture while using bucket-based scheduling to enhance system performance.
    
    \textbf{Attention Mechanism.} This type of technology focuses on reducing memory and computational bottlenecks in the attention module. vLLM’s PagedAttention\cite{kwon2023efficient} organizes KV cache into fixed-size blocks (similar to OS virtual memory), reducing fragmentation from 35\% to 5\% and boosting throughput by 2.4×. FlashAttention\cite{dao2022flashattention} fuses attention operations into a single kernel, accelerating computation by 1.8× on A100 GPUs through tiling and SRAM-based intermediate storage. FlashAttention-2\cite{dao2023flashattention2} further improves parallelization and supports advanced features like multi-query attention. Grouped Query Attention \cite{shazeer2019fasttransformerdecodingwritehead} balances efficiency and quality by grouping attention heads, enabling scalable context windows and batch sizes while retaining near-MHA performance. However, these work do not focus on requests scheduling.
    
    \textbf{Dynamic Batching.} It aims to maximize GPU utilization by adapting to variable-length sequences while currently limited work consider batching dynamically. Orca\cite{Orca280922} employs iteration-level scheduling, where batch sizes are determined per forward-backward pass. By replacing completed sequences immediately, it achieves higher GPU utilization than static batching, however, it only considers simple P/D disaggregation. FastServe \cite{wu2024fastdistributedinferenceserving} uses preemptive scheduling with a skip-join Multi-Level Feedback Queue to minimize JCT. Leveraging the semi-stateless nature of LLM inference, it dynamically prioritizes jobs based on input length, skipping lower-priority queues to reduce latency. FastServe also integrates efficient GPU memory management to handle variable workloads while the batch algorithm is not adaptive to dynamic workloads.
    
     \section{Conclusions and Future Work}
    \label{sec:conclusion}
    In this paper, we present BucketServe, a bucket-based dynamic batching framework designed to address the challenges of efficient LLM inference serving under concurrent and heterogeneous workloads. BucketServe introduces an adaptive bucketing mechanism that groups requests by sequence length, thereby reducing padding overhead and improving GPU memory utilization. The system further employs a dynamic batching controller that adjusts batch sizes in real time based on current GPU memory constraints, ensuring memory safety and maximizing throughput. Our comprehensive evaluations demonstrate that BucketServe achieves substantial improvements over existing baselines, including up to 3.58× higher throughput and nearly 2× greater system load capacity, while maintaining SLO attainment and incurring minimal overhead. These results validate the effectiveness of combining adaptive bucketing with dynamic batching for scalable, high-performance LLM serving. As for future work, we would like to explore multi-level load balancing strategies that can further enhance the system's performance under varying workloads on multi-nodes clusters.
   
  \section*{Acknowledgments}
   This work is supported by Guangdong Basic and Applied Basic Research Foundation (No. 2024A1515010251, 2023B1515130002), Guangdong Special Support Plan (No. 2021TQ06X990), Key Research and Development and Technology Transfer Program of Inner Mongolia Autonomous Region (2025YFHH0110), Shenzhen Basic Research Program under grants JCYJ20220818101610023 and JCYJ20240809180935001.

\bibliographystyle{IEEEtran}
\bibliography{ref}

\end{document}